\documentclass[a4paper,11pt,titlepage]{article}\usepackage[]{graphicx}\usepackage[]{xcolor}
\makeatletter
\def\maxwidth{ \ifdim\Gin@nat@width>\linewidth
    \linewidth
  \else
    \Gin@nat@width
  \fi
}
\makeatother

\definecolor{fgcolor}{rgb}{0.345, 0.345, 0.345}

\usepackage{framed}


\definecolor{shadecolor}{rgb}{.97, .97, .97}
\definecolor{messagecolor}{rgb}{0, 0, 0}
\definecolor{warningcolor}{rgb}{1, 0, 1}
\definecolor{errorcolor}{rgb}{1, 0, 0}
\newenvironment{knitrout}{}{} 

\usepackage{alltt}
\usepackage[pagebackref]{hyperref}
\renewcommand*{\backref}[1]{}
\renewcommand*{\backrefalt}[4]{{[\ifcase #1 Not cited.\or Cited on page~#2.\else Cited on pages #2.\fi ]}}

\usepackage[round, authoryear]{natbib}
\usepackage{lmodern,ifthen,amsmath,amssymb,csquotes,mathtools,tikz,calc,fixmath,bm,doi,float,chngcntr,etoc}
\etocsettocstyle{}{}
\mathtoolsset{showonlyrefs}
\usepackage{graphicx,booktabs,subfig}
\usepackage[draft]{fixme}
\fxsetup{theme=color}

\usetikzlibrary{positioning,calc,decorations.pathmorphing}

\def\beq{\begin{equation}}
\def\eeq{\end{equation}}

\def\edgecovsymbol{x}

\newcommand{\edgecov}[1][]{\edgecovsymbol\ifthenelse{\equal{#1}{}}{}{_{#1}}}

\def\netcovsymbol{z}
\newcommand{\netcov}[1][]{\netcovsymbol\ifthenelse{\equal{#1}{}}{}{_{#1}}}

\def\statsym{g}

\newcommand{\stat}[1][]{\statsym\ifthenelse{\equal{#1}{}}{}{_{#1}}}

\def\targetsym{t}
\newcommand{\target}[1][]{\targetsym\ifthenelse{\equal{#1}{}}{}{_{#1}}}

\def\paramsymbol{\theta}
\def\lparamsymbol{\beta}
\newcommand{\param}[1][]{\paramsymbol\ifthenelse{\equal{#1}{}}{}{_{#1}}}

\def\lparamv{\bm{\lparamsymbol}}
\newcommand{\lparam}[1][]{\lparamsymbol\ifthenelse{\equal{#1}{}}{}{_{#1}}}

\def\0{\bm{0}}

\def\nactors{n}

\def\sampidx{s}
\def\samp{S}

\DeclareMathOperator{\Var}{\mathbb{V}\kern-0.072em ar}
\DeclareMathOperator{\Cor}{\mathbb{C}\kern-0.072em or}
\DeclareMathOperator{\vecf}{vec}

\def\pval{\ensuremath{P\text{-val.}}}

\def\sigsym{{\tiny\ensuremath{\star}}}
\newcommand{\sigfw}[1]{\raisebox{0.5em}{\ifthenelse{\equal{#1}{0}}{\phantom{\sigsym\sigsym\sigsym}}{}\ifthenelse{\equal{#1}{1}}{\sigsym\phantom{\sigsym\sigsym}}{}\ifthenelse{\equal{#1}{2}}{\sigsym\sigsym\phantom{\sigsym}}{}\ifthenelse{\equal{#1}{3}}{\sigsym\sigsym\sigsym}{}}}
\newcommand{\sig}[1]{\raisebox{0.5em}{\ifthenelse{\equal{#1}{0}}{}{}\ifthenelse{\equal{#1}{1}}{\sigsym}{}\ifthenelse{\equal{#1}{2}}{\sigsym\sigsym}{}\ifthenelse{\equal{#1}{3}}{\sigsym\sigsym\sigsym}{}}}

\newcommand{\pkg}[1]{\texttt{#1}}

\newlength{\widthoffootnote}

\newcommand{\Model}[1]{\emph{Model~#1}}

\newcommand{\citepshort}[2][]{\nocite{#2}\citetext{\citetalias{#2},~\citeyear{#2}}\ifthenelse{\equal{#1}{}}{}{, #1}}
\newcommand{\citealpshort}[2][]{\nocite{#2}\citetalias{#2},~\citeyear{#2}\ifthenelse{\equal{#1}{}}{}{, #1}}

 \usepackage{etoc}
\usepackage{authblk}
\etocsettocstyle{}{}

\IfFileExists{upquote.sty}{\usepackage{upquote}}{}
\begin{document}
\thispagestyle{empty}
\pagenumbering{gobble}

\def\papertitle{A Tale of Two Datasets:\\ Representativeness and Generalisability of Inference for Samples of Networks}
\def\authorblock{\author[1]{Pavel N. Krivitsky\thanks{\href{mailto:p.krivitsky@unsw.edu.au}{\texttt{p.krivitsky@unsw.edu.au}}}}
\author[2]{Pietro Coletti}
\author[2,3]{Niel Hens}
\affil[1]{Department of Statistics and UNSW Data Science Hub

  School of Mathematics and Statistics

  University of New South Wales

  Sydney, Australia

~}
\affil[2]{I-BioStat

  Data Science Institute

  Hasselt University

  Hasselt, Belgium

~}
\affil[3]{Centre for Health Economics and Modelling Infectious Diseases

  Vaccine and Infectious Disease Institute

  University of Antwerp

  Antwerp, Belgium}
}

\renewcommand{\thefootnote}{\fnsymbol{footnote}}
\title{Rejoinder to Discussion of ``\papertitle''}
\authorblock
\date{}
\maketitle

\listoffixmes

\renewcommand{\thefootnote}{\arabic{footnote}}

\clearpage

\pagenumbering{arabic}

We thank Prof\@. Michael Stein for selecting our article for discussion and the four discussants for their praise and criticism alike. The discussants make a number of important points about both our particular analysis and the more general statistical and computational issues surrounding it, and we provide our thoughts on these in turn. It is our hope that this rejoinder will provide both additional guidance for practitioners and directions and suggestions for future research, both in methodology and in applications.

On some points, we can only concur. We see the link between simulation-based residuals and score tests drawn by \citet[Sec.~3]{ScFr23d} as a promising way to formalize the approach proposed in Section~4.3 of the article to diagnose between-network lack-of-fit and suggest additional predictors by regressing residuals on candidates. Our argument is heuristic and empirical, whereas a rigorous characterization of precisely which augmented ERGM constitutes the score test alternative hypothesis for a particular residual regression test would certainly guide the model selection better or suggest better tests. Similarly, their suggestion (Sec.~4) to implement user-friendly score testing is well taken, though care must be taken to ensure that partially observed scenarios are handled correctly.

\citet[Sec.~2]{Ni23d} makes a case for using family roles rather than age--gender categories as predictors of contacts and makes a point that some family compositions are categorically different from others, so the assumption of smooth network size effects implicit in our use of polynomials is questionable. Although we stand behind our model's adequacy based on the diagnostics we present, we agree in principle and had, in fact, considered the family role and categorical composition approach. Unfortunately, the uncertainty inherent in inferring roles and relations from demographics would create an errors-in-variables problem, greatly complicating the analysis. Thus, we reiterate the suggestion we made in the article: that future surveys collect this information directly.

Other points, we address in more detail below.

\section{Sample Size, Network Size, and Power}

\citet{Ve23d} points out that in determining the sample size requirements for inference for samples of networks, one must take into account both the number of networks to sample ($\samp$) and the sizes of the individual networks ($\nactors_\sampidx$). He also notes that in our application, the size distribution of networks in the population is well-defined and exogenous.

We would add that even so, the study designer still has some discretion in oversampling or stratifying on network sizes and compositions. The relative amounts of (Fisher) information contained in networks of different sizes in turn depends on the scaling regime of the model and can range from constant in $\nactors_\sampidx$ to linear to quadratic, depending on what network feature (edge count, mean degree, or density, respectively) is asymptotically invariant to $\nactors_\sampidx$ under the model \citep{KrKo15q}.

\begin{knitrout}
\definecolor{shadecolor}{rgb}{0.969, 0.969, 0.969}\color{fgcolor}\begin{figure}

{\centering \includegraphics[width=0.9\textwidth]{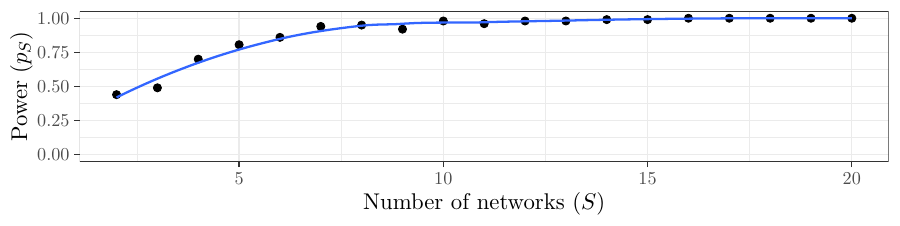} 

}

\caption{Empirical power $p_\samp$ for detecting a gender homophily effect with an assumed $\param_{\text{homophily}}=1.1$ as a function of number of networks $\samp$, where each network has size 8 nodes and 20 edges in the scenario of \citet[Sec.~1]{Ve23d}.}\label{fig:empirical-power}
\end{figure}

\end{knitrout}

In the simplified situation of networks of the same size and with a given edge count, the approach suggested by \citet{Ve23d}  can be used to infer the empirical power $p_\samp$ (as defined by \citeauthor{Ve23d}), which we illustrate in Figure~\ref{fig:empirical-power}. This is, of course, only a proof of concept, and should be adapted to the situation at hand.

\section{Multicollinearity and Variance Inflation Factors}

\citet{Ve23d} raises the question of multicollinearity and variance inflation in ERGMs for small networks. Outside of the extreme scenarios in which multicollinearity leads to outright nonidentifiability (i.e., Section~4.1 and Appendix~C in the article), whether it and high VIFs are a serious concern depends on how the parameter estimates are intended to be used \citep[for example]{O07c}, and we examine their impact on our analysis here.

In developing this discussion, we identified a calculation error that led to incorrect standard errors for the parameter estimates given in Table 1 in the article, with those related to 2-stars, triangles and their interactions with network size too small by an order of magnitude. None of the reported substantive conclusions about overall effects of 2-stars, triangles and network size change. We present the corrected standard errors in Table~\ref{tab:coef} and the corrected omnibus and contrast tests in Tables~\ref{tab:omnibus-tests} and~\ref{tab:contrast-tests}, respectively.

\begin{table}
  \caption{\label{tab:coef}Parameter estimates and corrected standard errors for \Model{1} and \Model{2}.}
  \footnotesize
  \begin{center}

\begin{tabular}{lrr}
\toprule
\multicolumn{1}{l}{Relationship Effect} & \multicolumn{2}{c}{Coefficient (S.E.)} \\
\cline{2-3}
\quad{}$\times$ Network-Level Effect & \Model{1}$^{\hphantom{\star\star\star}}$ & \Model{2}$^{\hphantom{\star\star\star}}$\\
\midrule
edges $\times$ $\log(\nactors_{\sampidx})$ & $-14.28 \; (3.78)^{\star\star\star}$ & $-13.78 \; (3.69)^{\star\star\star}$\\
\quad $\times$ $\log^2(\nactors_{\sampidx})$ & $5.69 \; (1.71)^{\star\star\star}$ & $5.47 \; (1.66)^{\star\star\phantom{\star}}$\\
\quad   if Brussels post code & $0.08 \; (0.19)^{\phantom{\star}\phantom{\star}\phantom{\star}}$ & $-0.02 \; (0.21)^{\phantom{\star}\phantom{\star}\phantom{\star}}$\\
\quad $\times$ $\log(\text{pop.\ dens.\ in post code})$ &  & $0.04 \; (0.03)^{\phantom{\star}\phantom{\star}\phantom{\star}}$\\
\quad   if on weekend & $0.14 \; (0.06)^{\star\phantom{\star}\phantom{\star}}$ & $0.13 \; (0.06)^{\star\phantom{\star}\phantom{\star}}$\\
2-stars & $1.91 \; (4.95)^{\phantom{\star}\phantom{\star}\phantom{\star}}$ & $1.14 \; (4.55)^{\phantom{\star}\phantom{\star}\phantom{\star}}$\\
\quad $\times$ $\log(\nactors_{\sampidx})$ & $-2.15 \; (5.87)^{\phantom{\star}\phantom{\star}\phantom{\star}}$ & $-1.22 \; (5.38)^{\phantom{\star}\phantom{\star}\phantom{\star}}$\\
\quad $\times$ $\log^2(\nactors_{\sampidx})$ & $0.34 \; (1.72)^{\phantom{\star}\phantom{\star}\phantom{\star}}$ & $0.07 \; (1.58)^{\phantom{\star}\phantom{\star}\phantom{\star}}$\\
triangles & $5.55 \; (11.55)^{\phantom{\star}\phantom{\star}\phantom{\star}}$ & $7.30 \; (10.72)^{\phantom{\star}\phantom{\star}\phantom{\star}}$\\
\quad $\times$ $\log(\nactors_{\sampidx})$ & $-3.46 \; (14.29)^{\phantom{\star}\phantom{\star}\phantom{\star}}$ & $-5.65 \; (13.28)^{\phantom{\star}\phantom{\star}\phantom{\star}}$\\
\quad $\times$ $\log^2(\nactors_{\sampidx})$ & $0.93 \; (4.38)^{\phantom{\star}\phantom{\star}\phantom{\star}}$ & $1.60 \; (4.09)^{\phantom{\star}\phantom{\star}\phantom{\star}}$\\
Young Child with Young Child & $8.60 \; (1.88)^{\star\star\star}$ & $8.66 \; (1.83)^{\star\star\star}$\\
Young Child with Preadolescent & $9.10 \; (1.88)^{\star\star\star}$ & $9.15 \; (1.84)^{\star\star\star}$\\
Preadolescent with Preadolescent & $8.17 \; (1.85)^{\star\star\star}$ & $8.24 \; (1.81)^{\star\star\star}$\\
Adolescent with Adolescent & $7.70 \; (1.84)^{\star\star\star}$ & $7.75 \; (1.80)^{\star\star\star}$\\
Young Child with Young Adult & $9.64 \; (2.07)^{\star\star\star}$ & $9.67 \; (2.05)^{\star\star\star}$\\
Preadolescent with Young Adult & $7.25 \; (1.85)^{\star\star\star}$ & $7.28 \; (1.81)^{\star\star\star}$\\
Adolescent with Young Adult & $7.73 \; (1.80)^{\star\star\star}$ & $7.82 \; (1.77)^{\star\star\star}$\\
Young Adult with Young Adult & $7.66 \; (1.85)^{\star\star\star}$ & $7.70 \; (1.81)^{\star\star\star}$\\
Young Child with Older Female Adult & $10.26 \; (1.85)^{\star\star\star}$ & $10.32 \; (1.81)^{\star\star\star}$\\
Preadolescent with Older Female Adult & $9.67 \; (1.85)^{\star\star\star}$ & $9.73 \; (1.80)^{\star\star\star}$\\
Adolescent with Older Female Adult & $8.90 \; (1.84)^{\star\star\star}$ & $8.96 \; (1.80)^{\star\star\star}$\\
Older Female Adult with Older Female Adult & $7.45 \; (1.87)^{\star\star\star}$ & $7.50 \; (1.83)^{\star\star\star}$\\
Young Child with Older Male Adult & $9.09 \; (1.87)^{\star\star\star}$ & $9.14 \; (1.82)^{\star\star\star}$\\
Preadolescent with Older Male Adult & $8.76 \; (1.83)^{\star\star\star}$ & $8.83 \; (1.79)^{\star\star\star}$\\
Adolescent with Older Male Adult & $8.20 \; (1.85)^{\star\star\star}$ & $8.26 \; (1.80)^{\star\star\star}$\\
Older Female Adult with Older Male Adult & $10.11 \; (1.84)^{\star\star\star}$ & $10.17 \; (1.80)^{\star\star\star}$\\
\quad   if child absent & $-1.22 \; (0.30)^{\star\star\star}$ & $-1.20 \; (0.30)^{\star\star\star}$\\
Older Male Adult with Older Male Adult & $6.59 \; (1.87)^{\star\star\star}$ & $6.66 \; (1.82)^{\star\star\star}$\\
Older Female Adult with Senior & $8.12 \; (1.82)^{\star\star\star}$ & $8.20 \; (1.78)^{\star\star\star}$\\
Older Male Adult with Senior & $7.51 \; (1.86)^{\star\star\star}$ & $7.58 \; (1.81)^{\star\star\star}$\\
Senior with Senior & $7.82 \; (1.81)^{\star\star\star}$ & $7.89 \; (1.77)^{\star\star\star}$\\
Adolescent with Young Child or Preadolescent & $8.07 \; (1.85)^{\star\star\star}$ & $8.13 \; (1.80)^{\star\star\star}$\\
Young Adult with Older Adult & $8.02 \; (1.84)^{\star\star\star}$ & $8.07 \; (1.80)^{\star\star\star}$\\
Young Child or Preadolescent with Senior & $8.29 \; (1.93)^{\star\star\star}$ & $8.34 \; (1.89)^{\star\star\star}$\\
Adolescent or Young Adult with Senior & $9.93 \; (2.09)^{\star\star\star}$ & $10.01 \; (2.08)^{\star\star\star}$\\
\bottomrule
\end{tabular}

Significance: $^{\star\star\star}\le 0.001<^{\star\star}\le 0.01< ^\star \le 0.05$
\end{center}
\end{table}

\begin{table}
  \caption{\label{tab:omnibus-tests} Selected omnibus tests for \Model{1}.}
  \footnotesize
  \begin{center}

\begin{tabular}{lrr}
\toprule
Effects & Wald $\chi^2$ (df) & $\pval$\\
\midrule
any 2-star & 20.9 (3) & \ensuremath{<0.001}\\
any triangle & 98.7 (3) & \ensuremath{<0.001}\\
any $\log(\nactors_\sampidx)$ or $\log^2(\nactors_\sampidx)$ & 78.9 (6) & \ensuremath{<0.001}\\
any $\log^2(\nactors_\sampidx)$ & 17.7 (3) & \ensuremath{0.001}\\
2-star or triangle $\log^2(\nactors_\sampidx)$ & 9.4 (2) & \ensuremath{0.009}\\
\bottomrule
\end{tabular}

\end{center}
\end{table}

\begin{table}
  \caption{\label{tab:contrast-tests} Selected contrasts for \Model{1}.}
  \footnotesize
  \begin{center}

\begin{tabular}{lrr}
\toprule
Contrast & Estimate (S.E.) & $\pval$\\
\midrule
Older Female vs. Male Adults with Young Children & $1.16 \; (0.47)$ & \ensuremath{0.013}\\
Older Female vs. Male Adults with Preadolescents & $0.91 \; (0.29)$ & \ensuremath{0.002}\\
Older Female vs. Male Adults with Adolescents & $0.70 \; (0.25)$ & \ensuremath{0.006}\\
Older Female $>$ Male Adults with Seniors (one-tailed) & $0.61 \; (0.31)$ & \ensuremath{0.026}\\
\bottomrule
\end{tabular}

\end{center}
\end{table}

A hurdle in using ERGM VIF in our specific case is that it is only well-defined in models with an edge count (i.e., intercept) effect  \citep{Du18d}. Our models do not use an intercept: our mixing effects cover all of the possible age--gender group combinations, visualized in Figure~4 in the article. (In the language of ANOVA, we use the \emph{means}, as opposed to the \emph{effects}, parametrization.)

\begin{figure}
  \begin{center}

\includegraphics[scale=0.8]{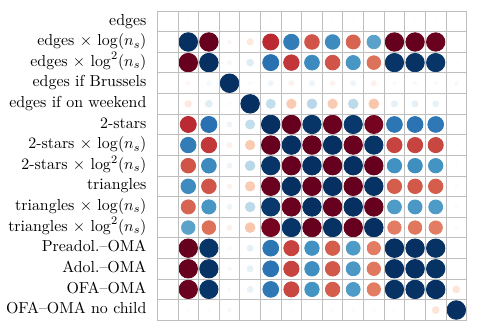}\includegraphics[scale=0.8]{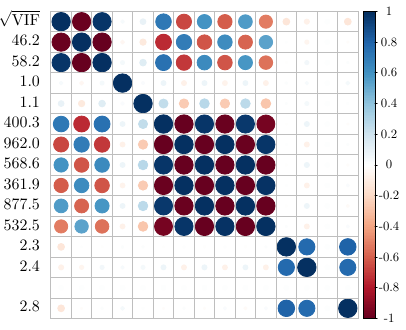}
\end{center}
\caption{\label{fig:corr-plots}Correlation matrices of selected parameter estimates for \Model{1} (left) and its effects reparametrization with root-variance-inflation-factors (right). Most mixing effects have been omitted for space and follow the same pattern as those shown.}
\end{figure}

We can nonetheless examine the impact of multicollinearity on our analysis directly. Figure~\ref{fig:corr-plots} (left) shows $\widehat{\Cor}(\vecf\hat{\lparamv})$ for a representative selection of parameters of \Model{1}. There are particularly strong correlations (positive and negative, following a ``checkerboard'' pattern across the polynomial network size effects) among the density (edges) and the endogenous (2-stars and triangles) effects and between those effects and the mixing effects. Among the mixing effects, all correlations are strong and positive. Referring back to Table~\ref{tab:coef}, as expected we observe particularly high standard errors for some of these endogenous effects; and if the VIF were well-defined, it would certainly be very high for all but a few of the parameter estimates.

However, this turns out to have little effect on the substantive conclusions. The endogenous effects are not individually significant in Table~\ref{tab:coef}, but when we perform Wald omnibus tests shown in Table~\ref{tab:omnibus-tests} (and Sec.~5.3 of the article), we find that the alternating signs of the coefficients combine with the ``checkerboard'' correlations to produce highly significant $\chi^2$ statistics. Then, when we visualize the network size effects, we take a linear combination of the estimates with positive coefficients---and the negative correlations result in the predicted effect for a given network size being far more precise than individual parameter estimates (Figure~\ref{fig:nseff2}).

\begin{knitrout}
\definecolor{shadecolor}{rgb}{0.969, 0.969, 0.969}\color{fgcolor}\begin{figure}

{\centering \subfloat[Edge (relative to $\nactors_\sampidx=2$)\label{fig:nseff2-1}]{\includegraphics[width=0.33\textwidth]{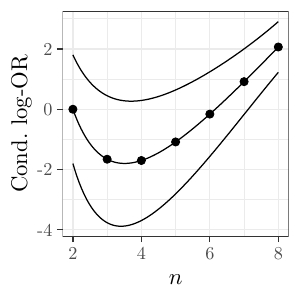} }
\subfloat[2-star\label{fig:nseff2-2}]{\includegraphics[width=0.33\textwidth]{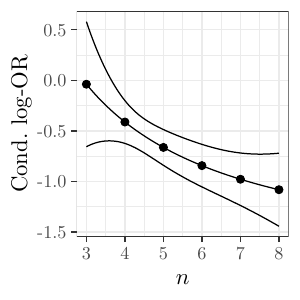} }
\subfloat[Triangle\label{fig:nseff2-3}]{\includegraphics[width=0.33\textwidth]{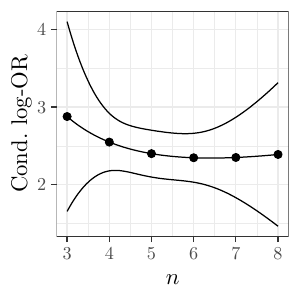} }

}

\caption[Estimated effects of network size on conditional log-odds of an instance of a graph feature with 1-standard-error bounds]{Estimated effects of network size on conditional log-odds of an instance of a graph feature with 1-standard-error bounds.}\label{fig:nseff2}
\end{figure}

\end{knitrout}

Lastly, but perhaps most importantly, when we compare different mixing cells by taking contrasts, a strong positive correlation between these estimates means that the contrasts in Table~\ref{tab:contrast-tests} are more precise than their constituents in Table~\ref{tab:coef}.

To examine the VIFs themselves, we reparametrize \Model{1} to use an edge count statistic and omit one of the mixing cells (the baseline), specifically, the Older Female Adult -- Older Male Adult cell, because it has the largest sample size. Figure~\ref{fig:corr-plots} (right) shows the correlations and the root-VIFs that result. We see that the intercept has ``absorbed'' the correlation between the mixing effects and the network size and endogenous effects, as well as some of the correlation among the mixing effects. Thus, VIFs are extremely high for the endogenous and network-size effects but modest for the mixing effects. However, it is important to keep in mind that those latter VIFs are for very specific contrasts---those with the baseline cell---which are not of substantive interest.

\section{Model Selection and Dyad-Dependent Effects}

\citet[Sec.~1]{Ni23d} highlights an important problem of efficient model selection for ERGMs. In our application in particular, most of the effects in the model were determined by substantive and inferential considerations, with only a few selected using diagnostics and AIC. More generally, however, the many possible relational effects representable by increasingly rich network data combined with the network-level effects on those effects can create a combinatorial explosion of possibilities.

Then, indeed, starting with dyad-independent effects and only adding dyad-dependent effects at the end as needed may turn out to be the only computationally feasible approach for many scenarios; but this approach is not without caveats. Model selection techniques such as stepwise regression can be sensitive to the order in which effects are considered for inclusion. In the case of ERGMs, theoretical and empirical studies are needed to confirm whether or not a model selection procedure that first selects dyad-independent effects and then selects dyad-dependent effects would produce systematically different results from one that selected dyad-dependent effects first or alongside. A corollary for our application would be that say, child effect choice may have been different if we had used residuals from a dyad-independent ``\Model{0d}'' to select it.

It may be possible to achieve the best of both worlds by substituting log-pseudolikelihood \citep{StIk90p} for the log-likelihood in the fit criterion (AIC or BIC); but whether this will also lead to systematically different results compared to selection using the true likelihood is an open question as well. (A further caveat is that for partially observed networks, the correct pseudolikelihood may not be that much easier to compute than the likelihood itself; see, for example, remarks by \citet[Sec.~5]{Kr17u}.)

\section{Data for Between-household Relations}

Between--household contacts, discussed by \citet{ScFr23d}, present a distinct modeling challenge from data collection perspective. The demographic-level information collected in contact diaries (and egocentric surveys in general) typically does not enable unambiguous identification of contacted individuals. This necessitates an approach such as that of \citet{KrMo17i}, in which survey data are used to estimate the global network statistics of interest, which are then used to estimate ERGM parameters by sufficiency. However, while residing in the same household is the strongest form of propinquity, between-household contacts are likely to have a very strong geographic propinquity effect as well, which cannot be modeled unless the contacts' geographic information is collected.

Moreover, one mechanism for violation of the local dependence assumption is that the activities household members participate in affect the connections both between and within households. Future studies should therefore collect the necessary information about overlapping social circles or activities people engage in, and perhaps models of \citet{WaRo16s} could be of use.

\section{User Guidance}

\citet{Ni23d} lists a number of frameworks and software tools that address a similar scenario to \pkg{ergm.multi} and asks which should be used when. Here, we would draw a distinction between the choice of a \emph{statistical model} and the choice of a \emph{software package} implementing the model. As we note in Section~3.2 of the article, our framework is a fixed-effects special case of that of \citet{SlKo16m}. Among the fixed-effects work listed, our reading of the work of \citet{VeSl21e} and \citet{StSc19m} is that the classes of models they span are either equivalent to or are subclasses of ours, the latter perhaps of the curved case outlined in Appendix~B of the article. Thus, as of this writing, the questions the practitioner should ask are 1) whether the networks in the sample can be safely assumed to be independent, and if not, whether between-network models such as those of \citet{WaRo16s} need to be specified; and 2) whether the fixed-effects model is adequate or leaves unaccounted-for heterogeneity that calls for a mixed-effects model.

Statistical software packages, on the other hand, are ever-evolving and interdependent, and not all of the above-listed models come with user-friendly and flexible code \citep[e.g.,][]{SlKo16m}. Thus, the best advice we can offer to the practitioner here is to research the workshops, the tutorials, and the vignettes provided with the packages and on their web sites, as well as recent papers that apply them, because those sources provide a continually updated view of a package's capabilities.

On a more positive note, this interdependence can make capabilities ``emerge'' without additional developer effort: for example, package \pkg{Bergm} for Bayesian estimation of ERGMs \citep{CaBo22s} does not have facilities for samples of networks, but it uses \pkg{ergm} \citep{KrHu23e} as its back-end and so can use all ERGM specifications implemented for \pkg{ergm}---including those provided by \pkg{ergm.multi} for modeling samples of networks. Thus, their combination may already be able to fit at least some special cases of \citet{SlKo16m}. 

Lastly, we concur that when demonstrating diagnostic techniques, it is important to demonstrate both fit and lack-of-fit look like. Appendix F of our article provides a sampling of those, but tutorials and similar can and should be developed as well.

\bibliography{bibliography}
\addcontentsline{toc}{section}{References}

\end{document}